\documentstyle[emulateapj,psfig]{article}

\newcommand{\be}{\begin{equation}}
\newcommand{\ee}{\end{equation}}
\newcommand{\ba}{\begin{eqnarray}}
\newcommand{\ea}{\end{eqnarray}}
\newcommand{\siml}{\lower4pt \hbox{$\buildrel < \over \sim$}}
\newcommand{\simg}{\lower4pt \hbox{$\buildrel > \over \sim$}}

\slugcomment{Accepted for publication in The Astrophysical Journal Letters}

\begin{document}

\title{GCRT J1745-3009 as a Transient White Dwarf Pulsar}

\author{Bing Zhang$^{1}$ and Janusz Gil$^{2,1}$}

\affil{$^1$ Department of Physics, University of Nevada, Las Vegas, NV
89154 \\ 
$^2$ Institute of Astronomy, Zielona G\'ora University,
Lubuska 2, 65-265 Zielona G\'ora, Poland}

\begin{abstract}
A transient radio source in the direction of the Galactic
Center, GCRT J1745-3009, exhibited 5 peculiar consecutive
outbursts at 0.33 GHz with the period of $77.13$ minutes and the
duration of $\sim 10$ minutes for each outburst. 
It has been claimed to be the prototype of a hitherto unknown
class of transient radio sources. We interpret it as a transient white
dwarf pulsar with a period of $77.13$ minutes. The $\sim
10$-minute flaring duration corresponds to the epoch when the radio
beam sweeps our line of sight. The bursting epoch corresponds to the
episodes when stronger sunspot-like magnetic fields
emerge into the white dwarf polar cap region during which the 
pair production condition is satisfied and the white dwarf
behaves like a radio pulsar. It switches off as the pair production
condition breaks down.
\end{abstract}

\keywords{stars: pulsars - stars: white dwarf -
radiation mechanism: coherent - radio}

\section{Introduction}
An enigmatic radio bursting source, GCRT J1745-3009, was
discovered recently in the direction of the Galactic
Center (Hyman et al. 2005). This source exhibited 5 peculiar consecutive
outbursts at 0.33 GHz with a period of $77.13$ minutes and a
duration of $\sim 10$ minutes for each outburst. The radiation is
very likely coherent as long as the distance is larger than 70 pc.
Although many efforts have been made to interpret it (Hyman et
al. 2005; Kulkarni \& Phinney 2005; Zhu \& Xu 2005; Turolla et
al. 2005), this behavior is hard to
understand in a straightforward way within the framework of known
astrophysical objects.  This source has been claimed to be the
prototype of a hitherto unknown class of transient radio
sources (Hyman et al. 2005). Here we show that the phenomenon is
naturally understood if GCRT J1745-3009 is a strongly magnetized
white dwarf, whose dipolar magnetic field defines a ``light
house'' beam from the polar cap region, just like what happens in
a radio pulsar.  The $77.13$-minute cycle is the rotation period
of the white dwarf, and the $\sim 10$-minute flaring duration
corresponds to the epoch when the radio beam sweeps our line of
sight. The bursting epoch corresponds to the episodes when the
pair production condition is satisfied, so that the white dwarf
can behave like a radio pulsar. When the pair production is turned
off, the flare ceases and the white dwarf enters its dormant
state. 

\section{White dwarf pulsars}

White dwarfs are intermediate compact objects that bridge normal main
sequence stars and more compact neutron stars. When our Sun collapses
to a white dwarf, the radius shrinks by a factor $\sim
100$. Conserving angular momentum gives a white dwarf period of a few
minutes. In reality, the observed white dwarf rotation period is
longer, typically hours to days (Kawaler 2004; Wickramasinghe \&
Ferrario 2000). Our suggested period $\sim 77$ min falls into the
lower end of the distribution, which is consistent with the fact that
this is the first one that was detected, since it takes a longer time
to identify the periodicity of 
the ones with longer periods, and since a shorter period favors pair
production which is the condition for coherent radio emission.
Conserving magnetic flux in the Sun during the collapse gives a
dipolar magnetic field of only $\sim 10^4$G at the white dwarf surface.
However, in reality there is a group of magnetized white dwarfs that
have a surface magnetic field in the range of $10^6 - 10^9$
G (Wickramasinghe \& Ferrario 2000). Some of them spin rapidly with
periods around an hour, which could be explained in terms of binary
evolution (Ferrario et al. 1997). These fast-rotating magnetized white
dwarfs are the objects we propose here to interpret the pulsating
behavior of GCRT J1745-3009.

Magnetic white dwarfs can mimic pulsars in various aspects. In
particular, it is well known that for a rotating, strongly-magnetized
object, the electromagnetic force dominates gravity and
thermal forces, and the natural outcome is a corotating
charge-separated magnetosphere (Goldreich \& Julian 1969).
Because of the unipolar effect, a large potential drop
develops across the polar cap region defined by the last open field
lines (Ruderman \& Sutherland 1975). The magnetized white dwarf idea
has been adopted to interpret the anomalous X-ray
pulsars (Pacz\'ynski 1990; Usov 1993, cf. Hulleman et al. 2000).
Below we show that the magnetized white dwarf model gives a
straightforward interpretation to the observational data of GCRT
J1745-3009. 

A period of $P \sim 77$ min defines a
light cylinder radius $R_{lc} = cP/2 \pi = 2.2 \times 10^{13} ~{\rm
cm} ~ (P/77~{\rm min})$. Given a typical white dwarf radius $R_{\rm
WD} = 5\times 10^8$ cm, the polar cap radius is
\begin{equation}
R_{pc} = R_{\rm WD} \left( \frac{R_{\rm WD}}{R_{lc}}\right)^{1/2}
= 2.4 \times 10^6 ~{\rm cm} R_{\rm WD,8.7}^{3/2}
\left(\frac{P}{77~{\rm min}}\right)^{-1/2}.
\end{equation}
Hereafter the convention $Q_x =(Q / 10^x)$ is adopted in cgs units.
Lacking a measurement of
the period derivative $\dot P$, one can not reliably estimate the spin
down rate and the dipolar surface magnetic field at the magnetic pole,
$B_p$. In the following we assume $B_p = 10^9$ G. The maximum
available unipolar potential drop across the polar cap reads
\begin{equation}
\Phi_{\rm max} = \frac{2 \pi^2 B_p R_{\rm WD}^3}{c^2 P^2}
=3.9 \times 10^{10} ~{\rm V} ~ B_{p,9} {R^3_{\rm WD,8.7}}
\left(\frac{P}{77~{\rm min}}\right)^{-2} ~.
\end{equation}
The maximum energy of the electrons accelerated in this potential drop
is $\gamma_{e,M} = e \Phi_{\rm max} / m c^2 = 7.6 \times 10^4 B_{p,9}
{R^3_{\rm WD,8.7}} ({P}/{77~{\rm min}})^{-2}$.
The potential drop is about 2 orders of magnitude smaller than that in
radio pulsars. 

The surface layer of a white dwarf is composed of a non-degenerate
electron gas and possibly an ionic lattice (Shapiro \& Teukolsky
1983). The lattice melting temperature is $T_m \sim 8.8 \times 10^5
~{\rm K}~ (\rho / 10^2 ~{\rm ergs ~ cm^{-3}})^{1/3} (Z/12)^{5/3}$,
where $\rho$ is the density in the surface layer, and $Z$ is the
atomic number (Mestal \& Ruderman 1967). Given a 
typical surface temperature $T_s \sim 3 \times 10^4$ K, we can see
that generally the surface is in the ionic lattice state. For an
anti-parallel rotator, i.e. ${\bf \Omega \cdot B} < 0$, where ${\bf
\Omega}$ and ${\bf B}$ are the vectors for the rotational and magnetic
axes, the polar cap region is populated with positively charged
particles. In a co-rotating magnetic white
dwarf magnetosphere, whether or not the surface can provide a free
ionic flow into the polar cap region depends on two factors, i.e,
whether thermionic emission could overcome the atomic cohesive
energy in the surface layer, and whether the free atoms are adequately
ionized. In strong magnetic fields, the ion cohesive energy is
enhanced and depends on the strength of the field, i.e. $\propto
B^{0.7}$ and is about several hundred eV for $B=10^{12}$ G for
iron (Jones 1986; Usov \& Melrose 1996). When $B$ is lower than $\sim
10^9$ G, which is the general case of the white dwarf pulsar discussed
here, however, the atoms are essentially not influenced by the magnetic
field (Lai 2001), so that the atom cohesive energies are similar to
the $B=0$ case. For carbon, which is likely the composition in the WD
surface layer, the cohesive energy is $\Delta\epsilon_c \sim 8$
eV/atom, and the first ionization energy of carbon is
$\Delta\epsilon_i \sim 11.3$ eV. The Goldreich-Julian (1969) charge
number density at the magnetic pole is $n_{_{\rm
GJ}}=(B_p/Pce)=1.5 \times 10^4 ~{\rm cm^{-3}}~ B_{p,9} (P/77~{\rm
min})^{-1}$ for GCRT J1745-3009, which is much lower than the number
density of carbon atoms in the surface 
layer, i.e. $n_{\rm atom} \sim \rho/Am_p = 5.0 \times 10^{24} ~{\rm
cm^{-3}} \rho_2$. For photoionization and thermionic ejection of the
ions, even if the ionization/ejection rate decreases exponentially
with temperature, the critical temperatures should be still several
tens smaller than the temperature defined by the cohesive/ionization
energy. Similar to the treatment for the neutron star
surface (Ruderman \& Sutherland 1975; Usov \& Melrose 1996), for the 
parameters of GCRT J1745-3009, this reduction factor is $\sim \log [Z e
\rho (kT)^{1/2} (Am_p)^{-3/2} P/B]=\log [3.8 \times 10^{16} (kT/2.6
~{\rm eV})^{1/2} (P/77 ~{\rm min}) B_{p,9}^{-1} \rho_2 (Z/6)
(A/12)^{-3/2} \sim 38$. As a result, the critical temperatures for
ionization of the carbon atom and for thermionic ejection of the
carbon atoms are $\sim 3.2\times 10^3$ K and $\sim 2.3\times 10^3$ K,
respectively, both are $\ll T_s \sim 3\times 10^4$ K, the typical
temperature of the white dwarf surface. The temperature
at the magnetic pole could be cooler (similar to Sunspots), but as
long as the temperature is higher than $\sim 3\times 10^3$ K, the
white dwarf surface is able to provide copious ions to supply a
Goldreich-Julian space charge limited flow from the polar cap
region (Arons \& Scharlemann 1979; Harding \& Muslimov 1998).
A similar conclusion applies for other surface
compositions, as well as for the case of a parallel rotator (${\bf
\Omega \cdot B} > 0$) in which case an electron free flow is supplied.

In a space-charge-limited flow, a charge depleted region is developed
in the polar cap region due to the general relativistic frame dragging
effect (Muslimov \& Tsygan 1992) and the curvature effect of the
magnetic field lines (Arons \& Scharlemann 1979). The ratio between
the potential drops developed for these two components is
roughly (Harding \& Muslimov 1998) $\sim (\kappa / \theta_{pc})~ 
{\rm ctan} \chi$, where $\chi$ is the inclination angle between the
magnetic and the rotational axes, $\theta_{pc}$ is the opening angle
of the polar cap region, $\kappa \sim (R_g/R_*)$, $R_g$ is the
Schwarzschild radius, and $R_*$ is the radius of the star (neutron
star or white dwarf).  For neutron star pulsars, the frame-dragging
term dominates unless the inclination is near 90 degrees (Harding \&
Muslimov 1998). For the white dwarf pulsars, with the parameters
of GCRT J1745-3009, we find that $\kappa \sim \theta_{pc} \sim
10^{-3}$, which means that both contributions are comparable. The
potential drop develops with height $h$ in the form of (Harding \&
Muslimov 1998) $\Phi (h) \sim (2 \pi B_p/Pc) R_{pc}^2 (h/R_{\rm
WD})^2$, which achieves the maximum potential $\Phi_{\rm max}$
essentially at $R \sim R_{\rm WD}$.  Without pair production,
electrons in the acceleration region could gain an energy close to
$\gamma_{e,M}$. Below we take a typical electron Lorentz
factor $\gamma_e=5\times 10^4$. 

\section{Pair production condition}

Can the electron-positron plasma needed by pulsar activity be
generated in such white dwarfs? This would require generation of seed
gamma photons with energies well exceeding the electron rest energy.
The dipolar field curvature radius in the white dwarf magnetosphere is
very large, i.e. $\rho_{d}=(4/3)\sqrt{R R_{lc}}=1.4 \times 10^{11} ~{\rm
cm} ~ (R/R_{\rm WD})^{1/2} (P/77 ~{\rm min})^{1/2}$. Near the surface,
non-dipolar magnetic fields could develop, as is the
case of the Sun (sunspots) and presumably also the case of neutron
stars (Ruderman \& Sutherland 1975; Arons \& Scharlemann 1979; Gil \&
Mitra 2001). However, even if we choose a much smaller
curvature radius, e.g. $\rho \sim 10^9$ cm, the typical curvature
radiation energy is still too small, i.e. $\epsilon_{\rm CR}=(3/2)
(\hbar c/\rho)\gamma_e^3 = 3.7 ~{\rm eV} ~ \gamma_{e,4.7}^3
\rho_9^{-1}$, well below the pair production threshold.

The typical thermal photon energy is $\epsilon_{\rm th} \sim 2.8 k T
= 7.3 T_{4.5}$ eV for $T_s \sim 3 \times 10^4$ K. Given the 
typical electron energy and the strength of the local magnetic field,
the resonant inverse Compton (IC) scattering is unimportant. The
typical resonant IC photon energy (Zhang et 
al. 1997) is $\epsilon_{\rm IC}^{R} \sim \gamma_{e} \epsilon_{\rm B} =
580 \gamma_{e,4.7} B_{p,9}$ keV, which is slightly larger than the
electron rest energy $m c^2 = 511$ keV, but does not meet the pair
production threshold. 

The most efficient gamma-ray production mechanism in a white dwarf
magnetosphere is non-resonant inverse Compton (IC) scattering (Zhang
et al. 1997). The typical gamma-ray photon energy reads 
\begin{eqnarray}
\epsilon_\gamma & = & \epsilon_{\rm IC}^{\rm NR}  =  {\rm min}
(\gamma_e^2 2.8 k T, \gamma_e m c^2) \nonumber \\
& = & {\rm min} (18 ~{\rm GeV} \gamma_{e,4.7}^2
T_{4.5}, 26 ~{\rm GeV} \gamma_{e,4.7}).
\label{photon-energy}
\end{eqnarray}
The second term takes into account the Klein-Nishina limit.
The mean free path for an electron to produce one IC gamma-ray photon
can be estimated as
\begin{equation}
l_{e} = (\sigma_{\rm IC} n_{ph})^{-1}=2.8 \times 10^9 ~{\rm cm} ~
T_{4.5}^{-3} \left( \frac{\sigma_{\rm IC}}{\sigma_{\rm T}}
\right)^{-1}~,
\label{le}
\end{equation}
where $\sigma_{\rm IC}$ and $\sigma_{\rm T}$ are the inverse Compton
cross section and Thompson cross section, respectively.
When a gamma-ray photon with a typical energy (\ref{photon-energy})
is emitted along the magnetic field line, it would interact with the
local magnetic fields to produce pairs (Sturrock 1971). The
mean free path for the photon to attenuate is (Ruderman \& Sutherland
1975) 
\begin{eqnarray}
l_{ph} & = & \chi \rho \left( \frac{B_q}{B_p} \right)
\left( \frac{2 m c^2}{\epsilon_\gamma} \right) \nonumber \\
& \simeq & {\rm max} (1.7 \times 10^8 ~{\rm cm} ~ \rho_9 B_{p,9}^{-1}
\gamma_{e,4.7}^{-2} T_{4.5}^{-1}, \nonumber \\
& & 1.2 \times 10^8 ~{\rm cm} ~
\rho_9 B_{p,9}^{-1} \gamma_{e,4.7}^{-1})~,
\label{lph}
\end{eqnarray}
where $\chi \sim 1/15$ has been used, $B_q=m_e^2c^3/e\hbar=4.4
\times 10^{13}$ G is the critical magnetic field, $m_e$, $e$, $c$, and
$\hbar$ are fundamental constants with their conventional
meanings. Equation (\ref{lph}) applies when $l_{ph} \ll R_{\rm WD}$ is
satisfied, so that the local magnetic field
does not decrease too much with respect to $B_p$. With the typical
parameters given above, $l_e$ is larger than $R_{\rm WD}$ and $l_{ph}$
is comparable to $R_{\rm WD}$. This means that the
condition for copious pair production is not satisfied (which is
defined by the condition $(l_e+l_{ph}) < R_{\rm WD}$), and the
white dwarf is below the so-called pair ``death line'' in the $P-B_p$
plane in analogy of radio pulsars (Ruderman \& Sutherland 1975; Zhang
et al. 2000; Hibschman \& Arons 2001; Harding \& Muslimov 2002;
Harding et al. 2002). This explains why GCRT J1745-3009 is dormant
under normal conditions, i.e. before and after the observed 5 bursting
cycles. 

A crucial point is that GCRT J1745-3009 lies not deep below the death
line, and can probably emerge out of the graveyard under some
circumstances. From Eqs. (\ref{le}) and (\ref{lph}), we can see that
$l_e$ sensitively depends on $T$, and $l_{ph}$ depends on both $\rho$ and
$B_p$. Since we have witnessed regularly enhanced sunspot activity
during the solar cycle, it would be natural to imagine
that more tangled magnetic structures would sometimes arise near the
white dwarf polar cap region. In fact, the star-spots analogous to
sunspots have most probably been observed in magnetized white
dwarfs (Maxted et al. 2000; Brinkworth et al. 2005). During such
magnetically active epochs, magnetic reconnection would heat up the
local magnetosphere (corona) to higher temperatures. Imagine during
one of these magnetic activities the temperature is raised by at least
a factor of 3, $l_e$ (Eq.[\ref{le}]) is greatly reduced to be (much)
shorter than $R_{\rm WD}$. Plenty of gamma-rays
(with typical energy 26 GeV) are generated. In the meantime, stronger
(say $B_p \sim ~{\rm several} \times 10^9$ G), more curved (say, $\rho
\sim 10^8$ cm) magnetic field structures emerge, so that $l_{ph}$ also
becomes much smaller than $R_{\rm WD}$. The IC $\gamma$-rays produced
by the primary electrons interact with local strong magnetic fields,
leading to copious production of electron-positron pairs. The white
dwarf pulsar is then ``turned on''.

According to the dipolar geometry, the emission altitude could be
estimated from the observed 10 min pulse duration, i.e.
\begin{eqnarray}
\frac{R_{e}}{R_{\rm WD}} & = & \left( \frac{10}{77} \frac{2\pi
 \sin \alpha} {3} \right)^{2} \frac{R_{lc}} {R_{\rm WD}} \nonumber \\
& \simeq & 3300 \left(\frac{P}{77~{\rm min}}\right) R_{\rm
WD,8.7}^{-1} (\sin \alpha)^{2}~,
\label{Re}
\end{eqnarray}
where $\alpha$ is the inclination angle between the
rotational axis and the magnetic axis. Surprisingly, this ``relative''
emission height falls nicely on the height-period correlation
discovered in radio pulsars (Fig.2 of Kijak \& Gil 2003), suggesting
some possible similar underlying physics.

After some time (5 periods for GCRT J1745-3009), the ``corona'' cools
down, and the strong magnetic fields disappear. Pair production is
turned off. The white dwarf returns back to its dormant phase. By
simple analogy, this time scale is different from the solar case.
This may be due to that the complex magnetic fields in white dwarfs
have a different origin than that in the Sun. These
magnetic fields may be arised from a 
convective surface layer of the white
dwarf (Steffen et al. 1995) or due to the Hall effect (Muslimov et
al. 1995). If the white dwarf accretes from a companion, one would
also expect large-scale effects on magnetic fields.

\section{Energetics \& detectability}

The spin down energy loss rate of GCRT J1745-3009
could be estimated as
\begin{eqnarray}
\dot E & = & \frac{(2 \pi)^4 B_p^2 R_{\rm WD}^6}{6 c^3 P^4} \nonumber
\\ 
& =& 3.3 \times 10^{26} ~{\rm erg~ s^{-1}}~ B_{p,9}^2 R_{\rm WD,8.7}^6
\left( \frac{P}{77 ~{\rm min}} \right)^{-4}~.
\end{eqnarray}
This is not a particularly energetic engine compared with normal
pulsars. The $0.33$ GHz radio flux at the flare is $\sim 1.67$ Jy. The
condition that the radio luminosity does not exceed the spindown
luminosity can be derived as 
\be
d < 0.8~{\rm kpc}~(\Delta \Omega_{-2})^{-1/2} B_{p,9} R_{\rm
WD,8.7}^3~,
\ee
where $\Delta \Omega$ is the unknown solid angle of the radio
emission, which for a slow rotator like GCRT J1745-3009, is
conceivable to be as small as 0.01. Given the uncertainties 
of $\Delta \Omega$, $B_p$ and $R_{\rm WD}$, the galactic center
distance ($\sim 8.5$ kpc) is not ruled out, although the source could
be much closer.

The maximum gamma-ray/X-ray flux would be
\begin{eqnarray}
F^{\gamma,X}_{\rm max} & = & \frac{\dot E}{\Delta\Omega D^2}
\nonumber \\
& = & 4.8 \times 10^{-17} ~{\rm erg~s^{-1}~cm^{-2}} \nonumber \\
& \times & (\Delta
\Omega)_{-2} B_{p,9}^2 R_{\rm
WD,8.7}^6 \left( \frac{P}{77 ~{\rm min}} \right)^{-4}
\left( \frac{D}{8.5~ {\rm kpc}} \right)^{-2}~.
\end{eqnarray}
This makes its gamma-ray/X-ray emission undetectable. The predicted
maximum X-ray flux is well consistent with the X-ray flux
upper limit $\sim 5 \times 10^{-10} ~{\rm
erg~s^{-1}~cm^{-2}}$ (Hyman et al. 2005).

If $B_p=10^9$ G, the expected spin down rate is 
\begin{equation}
\dot P = \frac{\dot E P^3}{4 \pi^2 I_{\rm WD}}
=8.2 \times 10^{-15} B_{p,9}^2 R_{\rm WD,8.7}^6 I_{\rm WD,50}^{-1}
\left( \frac{P}{77 ~{\rm min}} \right)^{-1}~,
\end{equation}
where $I_{\rm WD} \sim 10^{50} ~{\rm g~cm^2}$ is the moment of inertia
of the white dwarf. Even with long term monitoring, such a small
spindown rate is difficult to measure. 

The apparent optical/IR magnitude of a white
dwarf at a distance of 8.5 kpc is $\sim$ (27-30). Extinction would
further suppress the optical flux. Deep IR exposure with large
telescopes may lead to the discovery of the counterpart of GCRT
J1745-3009, especially if
the source is at a closer distance than that of the Galactic center.
Line features, if detected, would give a direct
measurement of the magnetic field strength through Zeeman spectroscopy
to test the hypothesis (Wickramasinghe \& Ferrario 2000).

\section{Discussion}
We have shown that the enigmatic transient radio source GCRT
J1745-3009 could be understood within the hypothesis that it is a
white dwarf pulsar. If this hypothesis is correct, the detection of
this powerful bursting radio source therefore suggests the discovery
of such a new type of pulsating, occasionally radio-loud, strongly
magnetized white dwarfs. The study of this object and future more
objects (if discovered) would also shed light on the
poorly understood coherent radio emission mechanism (e.g. Melrose
2004 for a review) of their brethren, neutron star pulsars.

Detecting such a transient white dwarf pulsar also suggests that some
``dead'' neutron star pulsars not deep below the death line may become
active again occasionally if strong sunspot-like magnetic fields
emerge into their polar cap regions. These transient radio pulsars are
awaiting being discovered.

\acknowledgements

We thank S. R. Kulkarni for drawing our attention to
this new phenomenon, D. Lai and J. Dyks for helpful discussion, and an
anonymous referee for helpful comments during the reviews.
This work is supported by NASA NNG04GD51G (BZ and JG) and by
grant 1 P03D 029 26 of the Polish State Committee for Scientific
Research (JG).

\end{document}